\documentclass[twocolumn,prd,aps,showpacs,secnumarabic]{revtex4}
\usepackage{graphicx}
\begin{document}

\title{Optical activity of neutrinos and antineutrinos}
\author{Ali Abbasabadi}
\affiliation{Department of Physical Sciences,
Ferris State University, Big Rapids, Michigan 49307, USA}
\author{Wayne W. Repko}
\affiliation{Department of Physics and Astronomy,
Michigan State University, East Lansing, Michigan 48824, USA}

\date{\today}
\begin{abstract}
\vspace*{0.5cm}
\hspace*{0.1cm}
Using the one-loop helicity amplitudes for low-energy $\nu\gamma\to\nu\gamma$ and $\bar\nu\gamma\to\bar\nu\gamma$ scattering in the standard model with massless neutrinos, we study the optical activity of a sea of neutrinos and antineutrinos. In particular, we estimate the values of the index of refraction and rotary power of this medium in the absence of dispersion.
\end{abstract}
\pacs{13.15.+g, 14.60.Lm, 14.70.Bh, 95.30.Cq}
\maketitle

\vskip1pc

\section{Introduction}
\label{sec:1}
The elastic neutrino-photon scattering process, $\nu\gamma\to\nu\gamma$, and the corresponding crossed channels have been the subject of numerous investigations related to astrophysical applications \cite{cm,mjl,ls,cy,dr93,liu,addr99,adr01,ar01}. When the center of mass energy $\sqrt{s}$ is much less than the $W$-boson mass $m_W$, the amplitudes for these processes have a relatively simple form. In particular, the forward helicity non-flip amplitudes for the elastic processes $\nu\gamma\to\nu\gamma$ and $\bar\nu\gamma\to\bar\nu\gamma$ with massless neutrinos can be expanded in powers of $s/m_W^2$ by making use of unitarity\cite{ar01}. The series are of leading order $s^2/m_W^4$, and the inclusion of higher powers of $s$, such as $s^3/m_W^6$, enables us to use the forward elastic scattering amplitudes for $\nu\gamma\to\nu\gamma$ and $\bar\nu\gamma\to\bar\nu\gamma$ to study the optical activity of a sea of neutrinos and antineutrinos. This property of a neutrino sea was recognized by Nieves and Pal \cite{np} and studied by Mohanty, Nieves, and Pal \cite{mnp} in the case where the photons move in a plasma and therefore experience dispersion. Dispersive effects give rise to mass dependent of terms of the form $k^2/m^2_e$  ($k^{\mu}$ denoting the photon momentum and $m_e$ the electron mass), which vanish in the vacuum. Here, we treat the vacuum case where $k^{\mu}$ satisfies $k^2=0$, and find a rather different dependence on $m_W$ and $m_e$ \cite{kn}.

In the next section, using the results of Ref.\,\cite{ar01}, we write the forward helicity amplitudes for the elastic scattering $\nu\gamma\to\nu\gamma$ and $\bar\nu\gamma\to\bar\nu\gamma$. We then present a calculation for the index of refraction and the rotary power for a sea of neutrinos and antineutrinos as a function of the energy of an incident photon, the temperature of the sea, and the neutrino degeneracy parameter $\xi \equiv \mu/T_\nu$, where $\mu$ is the chemical potential for neutrinos and $T_\nu$ is the temperature of the sea.

\section{Optical activity}
\label{sec:2}

Using the Eqs.\,(24)--(26) of the Ref.\,\cite{ar01}, the helicity non-flip amplitudes, ${{\cal A}^{\nu\gamma\to\nu\gamma}_{\lambda\lambda}}(s)$, for the forward elastic scattering of a photon with helicity $\lambda$ and an electron neutrino when $\sqrt{s} \ll m_W$, can be written as
\begin{eqnarray}
{{\cal A}^{\nu\gamma\to\nu\gamma}_{++}}(s) & = &
\frac{-\alpha^2s^2}{8m_W^4\sin^2\theta_W}\,h(s)\,,\label{nu1}\\\
{{\cal A}^{\nu\gamma\to\nu\gamma}_{--}}(s) & = &
\frac{-\alpha^2s^2}{8m_W^4\sin^2\theta_W}\,h(-s)\,,\label{nu2}
\end{eqnarray}
where
\begin{eqnarray} \label{h1}
h(s) & = & -4 -\frac{16}{3}\ln\left(\frac{m_W^2}{m_e^2}\right)
\nonumber\\
&&
+\frac{m_e^2}{m_W^2}\left[\left( \frac{8}{3}
-\frac{8}{3}\frac{s}{m_e^2} \right)
\ln\left(\frac{m_W^2}{m_e^2}\right)\right.\nonumber\\
&&
\left.
-10 +\frac{64}{9}\frac{s}{m_e^2}\right]\,.
\end{eqnarray}
Here, $\theta_W$ is the weak mixing angle, $\alpha$ is the fine structure
constant, and we have neglected higher powers of $m_e^2/m_W^2$
and $s/m_W^2$. Using the Eq.\,(12) of the Ref.\,\cite{ar01}, the
helicity non-flip amplitudes, ${{\cal A}^{\bar\nu\gamma\to\bar\nu\gamma}_{\lambda\lambda}}(s)$,
for forward elastic scattering of a photon of helicity $\lambda$
and an antineutrino, can be expressed as
\begin{eqnarray}
{{\cal A}^{\bar\nu\gamma\to\bar\nu\gamma}_
{\lambda \lambda}}(s) & = &
{{\cal A}^{\nu\gamma\to\nu\gamma}_
{-\lambda -\lambda}}(s)
\,,\label{cpt1}\
\end{eqnarray}
which is a consequence of $CPT$ invariance.

To calculate the optical activity of a neutrino-antineutrino sea,
we consider a photon of helicity $\lambda$ and energy $\omega$
traversing a bath of neutrinos and antineutrinos that are
in thermal equilibrium at the temperature $T_\nu$. To give an order of
magnitude estimate of the index of refraction $n_{\lambda}$ of this sea, we
write \cite{ll,gw}
\begin{eqnarray}\label{n1}
n_{\lambda} - 1 &=&
\frac{2\pi}{\omega^2}
\int dN_{\nu}\,f_{\lambda\lambda}^{\nu\gamma\to\nu\gamma}(0)\nonumber\\
&&
+\frac{2\pi}{\omega^2}
\int dN_{\bar\nu}\,f_{\lambda\lambda}^{\bar\nu\gamma\to\bar\nu\gamma}(0)
\,,
\end{eqnarray}
which is a generalization of the Eq.\,(57) of the
Ref.\,\cite{ar01}. The forward-scattering amplitudes
$f_{\lambda\lambda}^{\nu\gamma\to\nu\gamma}(0)$ and
$f_{\lambda\lambda}^{\bar\nu\gamma\to\bar\nu\gamma}(0)$, from the
Eq.\,(58) of the Ref.\,\cite{ar01}, are
\begin{eqnarray}
f_{\lambda\lambda}^{\nu\gamma\to\nu\gamma}(0) &=&
 \frac{\omega}{4\pi s}
{\cal A}^{\nu\gamma\to\nu\gamma}_{\lambda\lambda}(s)
\,,\label{fscat}\\
f_{\lambda\lambda}^{\bar\nu\gamma\to\bar\nu\gamma}(0) &=&
 \frac{\omega}{4\pi s}
{\cal A}^{\bar\nu\gamma\to\bar\nu\gamma}_{\lambda\lambda}(s)
\,,\label{fscatbar}
\end{eqnarray}
and $dN_\nu$ and $dN_{\bar\nu}$ are the usual Fermi-Dirac distributions,
\begin{eqnarray}
dN_{\nu} &=& \frac{1}{(2\pi)^3}
\frac{d^{3}\vec{p}_\nu}{e^{(E_{\nu}-\mu)/T_\nu} + 1}
\,,\label{dn}\\
dN_{\bar\nu} &=& \frac{1}{(2\pi)^3}
\frac{d^{3}\vec{p}_{\bar\nu}}{e^{(E_{\bar\nu}+\mu)/T_\nu} + 1}
\,.\label{dnbar}
\end{eqnarray}
Here, $T_\nu$ is the temperature of the neutrino-antineutrino sea,
$\mu$ is the chemical potential for the neutrinos \cite{p93},
$\vec{p}_\nu$ and $E_\nu$ are the momentum and energy of a
neutrino, and $s = 4\omega E_\nu\,
{\sin}^2(\theta_{\nu\gamma}/2)$, where $\theta_{\nu\gamma}$ is the
angle between the incoming photon and the incoming neutrino (with
similar definitions for the antineutrino). From
Eqs.\,(\ref{n1})--(\ref{fscatbar}) we obtain
\begin{eqnarray}\label{n2}
n_{\lambda} - 1 &=&
\int \frac{dN_{\nu}}{2\omega s}\,
{\cal A}^{\nu\gamma\to\nu\gamma}_{\lambda\lambda}(s)\nonumber\\
&&
+\int \frac{dN_{\bar\nu}}{2\omega s}\,
{\cal A}^{\bar\nu\gamma\to\bar\nu\gamma}_{\lambda\lambda}(s)
\,\,.
\end{eqnarray}

\begin{figure}[h]
\centering\includegraphics[width=2.75in]{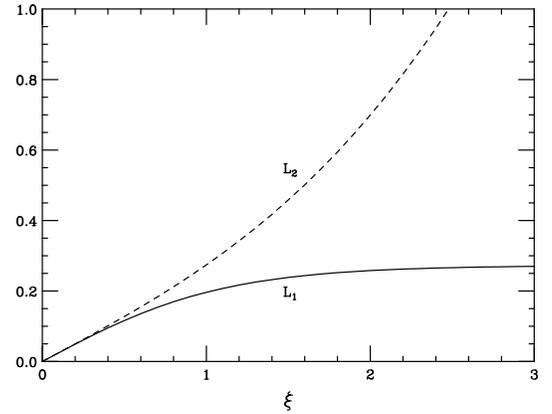}
\vspace{.10in}
\caption{The asymmetry parameters $L_1$ and $L_2$, as
functions of the neutrino degeneracy parameter
$\xi \equiv \mu/T_\nu$. The relations $L_1(-\xi) = -L_1(\xi)$
and $L_2(-\xi) = -L_2(\xi)$ hold.} \label{L1L2xi}
\end{figure}

Due to the apparent baryon asymmetry in the universe, it is
natural to consider the possibility of the lepton asymmetry. To
study this possibility, and for numerical calculations of the index of
refraction and the rotary power, it is convenient to introduce
the following ratios
\begin{eqnarray}
L_1 & \equiv &
\frac{3}{11}
\frac{N_\nu - N_{\bar\nu}}{N_\nu + N_{\bar\nu}}
\,,\label{ratioL1}\\
L_2 & \equiv &
\frac{2\pi^2}{11\zeta(3)}\frac{N_\nu - N_{\bar\nu}}{T_\nu^3}
\,,\label{ratioL2}\
\end{eqnarray}
where $N_\nu$ and $N_{\bar\nu}$ are the number densities for the
neutrino and antineutrino, respectively, and $\zeta(x)$ is
the Riemann zeta function. From the Eqs.\,(\ref{dn}), (\ref{dnbar}),
(\ref{ratioL1}), and (\ref{ratioL2}), we find
\begin{eqnarray}
L_1 & = &
\frac{1}{11}\frac{\xi^3 + \pi^2 \xi}
{I(\xi)\xi^3 + 2 \ln2 \,\xi^2 + 3\zeta(3)}
\,,\label{ratioL1a}\\
L_2 & = &
\frac{1}
{33\zeta(3)}\left(\xi^3 + \pi^2 \xi \right)\,,\label{ratioL2a}\
\end{eqnarray}
where
\begin{equation}
I(\xi) = \int_{0}^{1} (1 - x)^2
\frac{e^{\xi x} - 1}{e^{\xi x} + 1}\,dx
\,,\label{I}\\
\end{equation}
and $\xi \equiv \mu/T_\nu$ is the neutrino degeneracy parameter.
As it is clear from the Eqs.\,(\ref{ratioL1a}) -- (\ref{I}), the
parameters $L_1$ and $L_2$ are solely functions of $\xi$. These
neutrino asymmetry parameters $L_1$ and $L_2$ are defined such
that for $|\xi| \ll 1$ they coincide with the customary definition
of the neutrino asymmetry parameter $L_\nu$ \cite{lp} for a sea of
relic neutrino-antineutrino that decoupled from photons in the
early universe
\begin{eqnarray}
L_1 \simeq L_2 \simeq L_\nu & \equiv &
\frac{N_\nu - N_{\bar\nu}}{N_\gamma} \simeq
\frac{\pi^2 \xi}{33\zeta(3)}
\,.\label{ratio2}\
\end{eqnarray}
Here, $N_\gamma$ is the photon number density
\begin{equation}
N_{\gamma} = \frac{1}{(2\pi)^3}
\int \frac{2d^{3}\vec{p}_{\gamma}}
{e^{E_{\gamma}/T_\gamma} - 1} =
\frac{2\zeta (3)T_\gamma^3}{\pi ^2}
\,,\label{ngamma}\\
\end{equation}
and the relation $T_\nu/T_\gamma = (4/11)^{1/3}$, which
holds for the present day temperatures of the relic neutrinos
and photons, is used. When $\xi$ is large, $L_1$ approaches a finite value, while $L_2$ increases. In the Fig.\,\ref{L1L2xi} $L_1$ and $L_2$ are plotted as functions of $\xi$. As it is clear from this figure, for $\xi \gtrsim 3$, $L_1$ is close to its maximum value of $3/11$.

\begin{figure}[h]
\centering\includegraphics[width=2.75in]{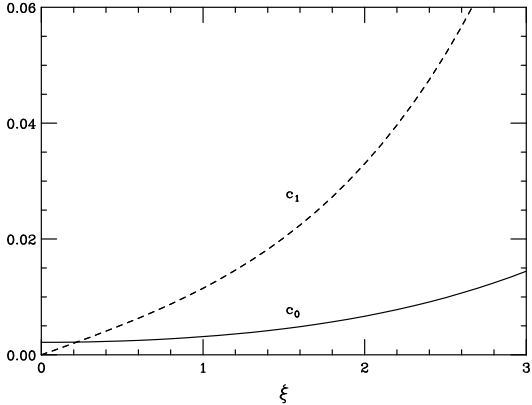}
\vspace{.10in}
\caption{The coefficients $c_0$ and $c_1$ as functions of the
neutrino degeneracy parameter
$\xi \equiv \mu/T_\nu$. The relations $c_0(-\xi) = c_0(\xi)$
and $c_1(-\xi) = -c_1(\xi)$ hold.} \label{c0c1xi}
\end{figure}

\begin{figure}[h]
\centering\includegraphics[width=2.75in]{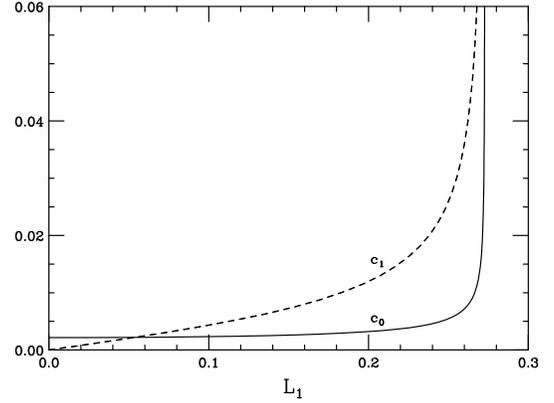}
\vspace{.10in}
\caption{The coefficients $c_0$ and $c_1$ as functions of the
asymmetry parameter $L_1$. The relations $c_0(-L_1) = c_0(L_1)$
and $c_1(-L_1) = -c_1(L_1)$ hold.} \label{c0c1L1}
\end{figure}

\begin{figure}[h]
\centering\includegraphics[width=2.75in]{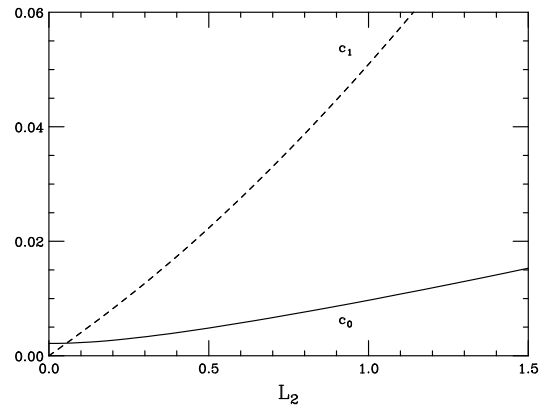}
\vspace{.10in}
\caption{The coefficients $c_0$ and $c_1$ as functions of the
asymmetry parameter $L_2$. The relations $c_0(-L_2) = c_0(L_2)$
and $c_1(-L_2) = -c_1(L_2)$ hold.} \label{c0c1L2}
\end{figure}

After using Eqs.\,(\ref{dn}) and
(\ref{dnbar}) for $dN_\nu$ and $dN_{\bar\nu}$,
Eqs.\,(\ref{nu1})--(\ref{cpt1}) for
the amplitudes
${\cal A}^{\nu\gamma\to\nu\gamma}_{\lambda\lambda}(s)$
and
${\cal A}^{\bar\nu\gamma\to\bar\nu\gamma}_{\lambda\lambda}(s)$,
and performing integrations in the Eq.\,(\ref{n2}), we obtain
\begin{eqnarray}
n_{+} - 1 &=& \frac{T_\nu^4}{m_W^4}\,c_0
+\frac{\omega T_\nu^5}{m_W^6}\,c_1\,,\label{np1}\\\
n_{-} - 1 &=& \frac{T_\nu^4}{m_W^4}\,c_0
-\frac{\omega T_\nu^5}{m_W^6}\,c_1\,,\label{nm1}
\end{eqnarray}
where (for $\alpha=1/137$)
\begin{eqnarray}
c_0  &=& \frac{\alpha^2}{3\pi^2 \sin^2\theta_W}
\left[\ln\left(\frac{m_W^2}{m_e^2}\right)
 + \frac{3}{4}\right]\nonumber\\
&& \times \left(\frac{1}{4}\,\xi^4 + \frac{\pi^2}{2}\,\xi^2
+ \frac{7\pi^4}{60}\right)\nonumber\\
&\simeq&
1.9 \times 10^{-4}\left(\frac{1}{4}\,\xi^4 + \frac{\pi^2}{2}\,\xi^2
+ \frac{7\pi^4}{60}\right),\;\label{c0}\\\
c_1  &=& \frac{4\alpha^2}{9\pi^2 \sin^2\theta_W}
\left[\ln\left(\frac{m_W^2}{m_e^2}\right)
 - \frac{8}{3}\right]\nonumber\\
&& \times \left(\frac{1}{5}\,\xi^5 + \frac{2\pi^2}{3}\,\xi^3
+ \frac{7\pi^4}{15}\,\xi\right)\nonumber\\
&\simeq&
2.2 \times 10^{-4}\left(\frac{1}{5}\,\xi^5 + \frac{2\pi^2}{3}\,\xi^3
+ \frac{7\pi^4}{15}\,\xi\right).\label{c1}
\end{eqnarray}
Notice that the $\xi$--dependent parts of the $L_2$, $c_0$, and
$c_1$ are related as
\begin{eqnarray}
\frac{1}{4}\frac{{\partial}}{\partial \xi}
\left(\frac{1}{5}\,\xi^5 + \frac{2\pi^2}{3}\,\xi^3
+ \frac{7\pi^4}{15}\,\xi\right)\;\;\;\;\;\;\; \hskip 30pt \nonumber\\
=\frac{1}{4}\,\xi^4 + \frac{\pi^2}{2}\,\xi^2
+ \frac{7\pi^4}{60}
\,\,,\label{c1c0}
\end{eqnarray}
\begin{equation}
\frac{{\partial}}{\partial \xi}
\left(\frac{1}{4}\,\xi^4 + \frac{\pi^2}{2}\,\xi^2
+ \frac{7\pi^4}{60}\right)
=\xi^3 + \pi^2 \xi \,\,.\label{c0L1}
\end{equation}
In the Figs.\,\ref{c0c1xi}--\ref{c0c1L2} the coefficients $c_0$ and $c_1$ are plotted as functions of $\xi$, $L_1$, and $L_2$, respectively.

The range of the validity of the Eqs.\,(\ref{np1})--(\ref{c1}), as far
as energy is concerned, is
related to that of the Eqs.\,(\ref{nu1})--(\ref{h1}), which is $s = 4\omega
E_\nu\, {\sin}^2(\theta_{\nu\gamma}/2)
 \ll m_W^2$. In the Eq.\,(\ref{n2}), if we change the
upper limits of the integrations on $E_\nu$ and $E_{\bar\nu}$ from
the infinity to $fT_\nu$, the contributions of these integrals to $c_0$
and $c_1$ in the Eqs.\,(\ref{np1}) and (\ref{nm1}) change, at most,
by 9\% and 18\%, respectively, if we use
$f = \sqrt{7^2 + \xi^2}$ (for $f = \sqrt{8^2 + \xi^2}$,
the corresponding changes are at most 5\% and 10\%). Here, we
set the following criterion:
\begin{equation}
4\omega fT_\nu  \ll \, m_W^2 \,,\label{criterion1}
\end{equation}
which for $f = \sqrt{7^2 + \xi^2}$ is
\begin{equation}
\omega T_\nu \sqrt{7^2 + \xi^2}
\ll 2\times10^{16} \;\;{\rm GeV}\cdot{\rm K}
\,,\label{criterion2}
\end{equation}
where $\omega$ is the photon energy in GeV, and $T_\nu$ is the neutrino
temperature in Kelvin.

From Eqs.\,(\ref{np1}) and (\ref{nm1}), we have
\begin{eqnarray}
n_{+} - n_{-} &=&
2\frac{\omega T_\nu^5}{m_W^6}\,c_1
\nonumber\\
&\simeq & 3.5 \times 10^{-77} c_1\,\omega T_\nu^5
\,,\label{npm1}
\end{eqnarray}
and the following approximate relation:
\begin{eqnarray}
n_{+} - 1 \simeq  n_{-} - 1 & \simeq &
\frac{T_\nu^4}{m_W^4}\,c_0
\nonumber\\
& \simeq & 1.3 \times 10^{-60} c_0 T_\nu^4
\,.\label{npm2}
\end{eqnarray}
Equation\,(\ref{npm2}) implies that the leading term of the index of refraction is independent of the helicity and the energy of the incident photon, as long as Eq.\,(\ref{criterion2}) is satisfied.

When linearly polarized light propagates through a medium that has different
indices of refraction for positive and negative helicities ($n_+ \not= n_-$),
the plane of polarization of the light rotates by an angle $\phi$, which is
\cite{b}
\begin{equation}
\phi = \frac{\pi}{\lambda_{\gamma}}\,(n_{+} - n_{-})\,l =
\frac{\omega}{2}\,(n_{+} - n_{-})\,l
\,,\label{phi1}
\end{equation}
where $\omega$ and $\lambda_{\gamma} = 2\pi/\omega$ are the energy
and wavelength of the photon and $l$ is the distance travelled by
photons in the medium. To estimate the specific rotary power,
$\phi/l$, for a sea of neutrinos and antineutrinos, we use
Eqs.\,(\ref{npm1}) and (\ref{phi1}) to obtain
\begin{eqnarray}
\frac{\phi}{l} &=&
\frac{\omega^2 T_\nu^5}{m_W^6}\,c_1
\nonumber\\
& \simeq &
8.9 \times 10^{-62}c_1\omega^2 T_\nu^5 \;\;\rm{rad/m} \,,\label{phi2}
\end{eqnarray}
where, again, $\omega$ is in GeV and $T_{\nu}$ is in Kelvin. A positive angle of rotation, $\phi > 0$, that is $n_+ > n_-$, corresponds to
a clockwise rotation (dextrorotation) of the plane of polarization of the
linearly polarized incident photons, as viewed by an observer that is
detecting the forward-scattered light. Thus, the optical activity of a
neutrino-antineutrino sea with $N_\nu > N_{\bar\nu}$  is that of a dextrorotary
medium. In addition, it is clear from
Eq.\,(\ref{phi2}) that the rotary power, $\phi/l$, varies as
$1/\lambda_{\gamma}^2$, which is the same as that of quartz and most
transparent substances for visible light.

\begin{figure}[h]
\centering\includegraphics[width=2.75in]{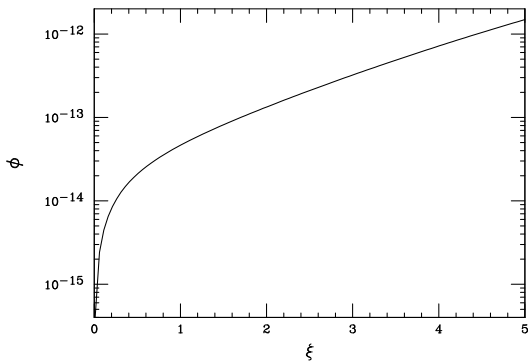}
\vspace{.10in}
\caption{The polarization rotation angle $\phi$ is shown as a function of $\xi$ for $\omega=10^{11}$\,GeV. The curve scales as $\omega^2$. \label{rot}}
\end{figure}

To get a rough estimate of rotation angle $\phi$ for linearly polarized photons propagating through the relic neutrino and antineutrino sea (with $\xi = 0.01$), we use Eq.\,(\ref{phi2}) with $l =ct$, $c = 3\times10^8$ m/s, $t \sim 15\times10^9$ years, $T_\nu \sim 2 $\,K, and $\omega \sim 10^{20}$ eV, and we find
\begin{equation}
\phi \sim 4\times10^{-16}\;\; {\rm rad}
\,,\label{phi3}
\end{equation}
which is exceedingly small. The dependence of $\phi$ on $\xi$ is shown in Fig.\,\ref{rot}. It should be noted that recent studies which include neutrino oscillation effects in the derivation of cosmological bounds on $\xi$ conclude that $|\xi|\lesssim 0.1$ \cite{Dolgov:2002ab,Abazajian:2002qx,Wong:2002fa}.
\section{Conclusions}
In conclusion, we note that using Eqs.\,(\ref{ratioL2}), (\ref{ratioL2a}),
(\ref{c1}), and (\ref{phi2}), for the neutrino degeneracy
parameter $|\xi| \ll 1$, the specific rotary power,
$\phi/l$, for a sea of neutrinos and antineutrinos is
\begin{equation}\label{phi4}
\frac{\phi}{l} =
\frac{112\pi G_F\alpha}{45\sqrt{2}}
\left[\ln\left(\frac{m_W^2}{m_e^2}\right)
- \frac{8}{3}\right]
\frac{\omega^2 T_\nu^2}{m_W^4}(N_\nu - N_{\bar\nu})
\,.
\end{equation}
This equation differs in several respects
from the corresponding result in Ref.\,\cite{mnp},
\begin{equation}\label{mnprot}
\frac{\phi}{l}=\frac{G_F\alpha}
{3\sqrt{2}\,\pi}(N_\nu-N_{\bar\nu})\frac{\omega_P^2}{m_e^2}\,,
\end{equation}
where, in the non-relativistic limit, the plasma frequency $\omega_P$ is
related to the electron number density $n_e$ as,
\begin{equation}
\omega_P^2=\frac{4\pi\alpha n_e}{m_e}\,.
\end{equation}
From Eqs.\,(\ref{phi4}) and (\ref{mnprot}), it is clear that the mass scales
for the case of photon dispersion versus no dispersion are very different and
this favors the dispersive regime. The size of specific rotary power in the
dispersive case is independent of the frequency of the propagating photons
and, apart from a dependence on the lepton asymmetry parameter
$L_\nu \propto N_\nu - N_{\bar\nu}$ shared by both expressions, is
controlled by the electron number density $n_e$. In addition, the
non-dispersive case depends on frequency and temperature.
These differences clearly indicate that
Eqs.\,(\ref{phi2}) and (\ref{mnprot}) represent complementary
limits in the treatment of the optical activity of a neutrino-antineutrino sea.

\acknowledgments We wish to thank Duane Dicus for numerous helpful
discussions. One of us (A.A.) wishes to thank the Department of Physics and
Astronomy at Michigan State University for its hospitality and computer
resources. This work was supported in part by the National Science Foundation
under Grant No. PHY-0070443.

\end{document}